\documentclass[amsmath,groupedaddress,showpacs,amssymb,aps,pra,preprint]{revtex4}
\usepackage{graphicx}
\usepackage{amsmath}

\newcommand{\figref}[1]{Fig.~\ref{fig:#1}}
\newcommand{\equref}[1]{Eq.~(\ref{eq:#1})}

\begin{document}
\title{Experimental demonstration of fractional orbital angular
momentum entanglement of two photons}
\author{S.S.R. Oemrawsingh}
\author{X. Ma}
\author{D. Voigt}
\author{A. Aiello}
\author{E.R. Eliel}
\author{G.W. 't Hooft}
\author{J.P. Woerdman}
\affiliation{Huygens Laboratory, Leiden University\\
P.O.\ Box 9504, 2300 RA Leiden, The Netherlands}
\begin{abstract}
The singular nature of a non-integer spiral phase plate allows easy
manipulation of spatial degrees of freedom of photon states. Using two such
devices, we have observed very high dimensional ($D > 3700$) spatial
entanglement of twin photons generated by spontaneous parametric
down-conversion.
\end{abstract}
\maketitle
In the early 1990s, it was realized that the quantum state of a photon in a
Laguerre-Gaussian laser mode LG$^l_p$ is an eigenstate of the orbital
angular momentum (OAM)
operator, with eigenvalue $l\hbar$ ($l$ integer) \cite{WOE92,NIE92}. The mode
functions LG$^l_p$ ($l=-\infty\dots\infty$, $p=0\dots\infty$) can be used as a
basis of an infinite-dimensional Hilbert space, and thus to describe, within
the paraxial approximation, a photon state represented by an arbitrary spatial
wave function.

According to theory, the twin photons created in spontaneous parametric
down-conversion (SPDC) are entangled in the full, infinite-dimensional
(spatial)
Hilbert space; in fact, they are maximally entangled \cite{ALL02,MON04}.
Several years ago, Mair~\emph{et~al.}\ \cite{ZEI01} demonstrated OAM
entanglement in a 3D subspace of that Hilbert space, in an experiment where
they used integer-OAM analyzers. It has been recently proposed that spatial
entanglement can be demonstrated in an \emph{infinite}-dimensional subspace
\cite{WOE04:3}.  This can be achieved by using \emph{fractional}-OAM
analyzers, in view of the fact that fractional OAM states are coherent
superpositions of an infinite number of LG$^l_p$ states with integer OAM.
Here we report an experimental implementation of this idea, employing two
half-integer OAM analyzers; we confirm the predicted entanglement of
fractional OAM and deduce a dimensionality of $D>3700$ per photon, the
entangled space having a dimensionality $D^2$.

\begin{figure}
\centerline{\includegraphics[height=3cm]{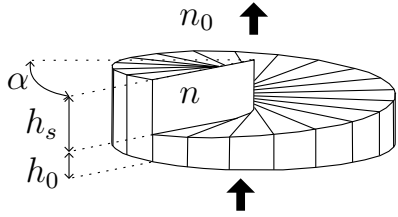}}
\caption{Schematic drawing of a spiral phase plate (SPP) with a step index
$\ell=h_s(n-n_0)/\lambda$, where $h_s$ is the step height, $n$ and $n_0$ are
the refractive indices of the SPP and the surrounding medium, respectively,
and $\lambda$ is the wavelength of the incident light.}\label{fig:spp}
\end{figure}
\begin{figure*}
\centerline{\includegraphics[height=4.481cm]{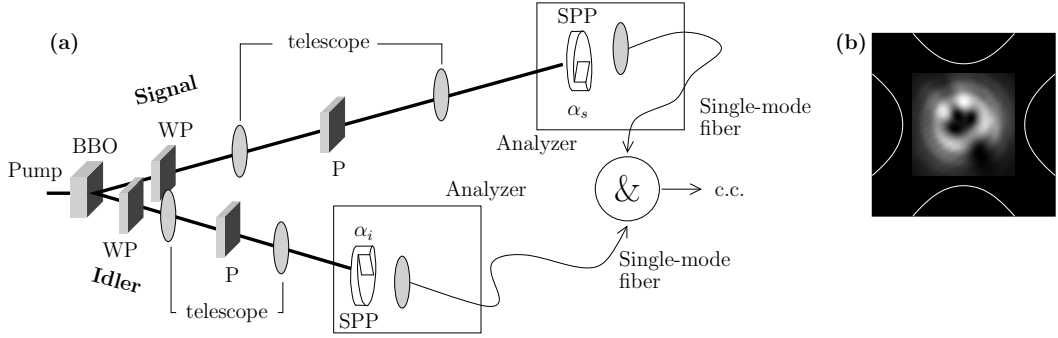}}
\caption{(a) The SPDC setup for establishing high-dimensional entanglement
with two analyzers. The SPDC crystal (BBO) is tilted so that the angle between
the signal and idler arms is $6^\circ$. In both arms, one encounters a
$0.5$~mm-thick wave plate ($\mathrm{WP}$) for walk-off compensation,  a
telescope that images the nonlinear crystal onto the SPP, a
Polarcor\texttrademark\ polarizer ($\mathrm{P}$) and a red filter (not shown)
to eliminate stray pump light.  The down-converted photons are focused with
microscope objectives onto single-mode fibers leading to
single-photon-counting modules; coincidence counts are measured. Both SPPs
can be rotated ($\alpha_s$ and $\alpha_i$) around the respective beam
axes. (b) An experimental far-field intensity pattern of the analyzer mode.
The white curves indicate the 90\% iso-intensity lines of the SPDC rings,
illustrating that the analyzer mode completely falls in the region of
twin-photon degeneracy.}\label{fig:quantumsetup}
\end{figure*}

The key element in our approach is the use of a spiral phase plate (SPP)
\cite{WOE94,WOE04}, shown in \figref{spp}, as part of our analyzer. This
plate imprints an azimuth-dependent phase retardation on an incident field. The
difference in phase retardation $\Delta\phi$ at azimuthal angles $0$ and
$2\pi$ is equal to $\Delta\phi\equiv 2\pi\ell$, whereby the plate adds an
average OAM of $\ell\hbar$ per photon. By adjusting the SPP parameters (see
caption of \figref{spp}), the
SPP index $\ell$ can be tuned to any desired value. If $\ell$ is
\emph{non-integer}, the output field is \emph{not} cylindrically symmetric
(cf.\ \figref{quantumsetup}(b)) and carries a mixed screw-edge dislocation in the near
field, i.e.\ a singularity with a transverse component; this dislocation is
associated with the radial edge of the SPP \cite{WOE04}. Thus, for all
non-integer $\ell$, the SPP is characterized by two parameters, namely the
step index $\ell$ itself and the orientation of the radial edge of the SPP,
$\alpha$.  The latter is
varied by rotating the SPP around the beam axis.

Our approach involves the use of two fractional-OAM analyzers (one per arm) in
an SPDC setup. Such an analyzer consists of a non-integer SPP, a single-mode
fiber and
a detector; the orientations of the radial edges of the SPPs are varied
(see \figref{quantumsetup}(a) below). Because the SPPs have a non-integer step
index $\ell$, the fractional OAM states provide a natural basis to describe
the two-photon state \cite{WOE04:3}. Each element of this fractional basis
can be written as an $\ell$ and $\alpha$ dependent superposition of an infinite
number of LG$^l_p$ states, extending over both $l$ and $p$. Because this
superposition changes when we reorient the SPP, we explore an
\emph{infinite}-dimensional subspace of the full, (spatial) Hilbert space by
systematically varying $\alpha$ (at constant step index $\ell$). We stress
that the contribution of LG$^l_p$ modes with $p\ne 0$ to this superposition is
significant; this is caused by the fact that the dislocation imprinted by the
SPP on the incident field has a mixed screw-edge character \cite{WOE04}.

\begin{figure*}
\centerline{\includegraphics[width=14.5cm]{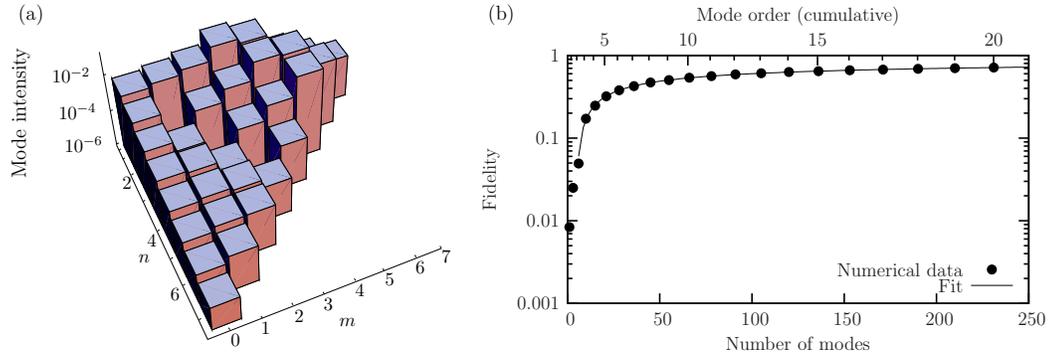}}
\caption{Decomposition of the analyzer mode, associated with an SPP with
$\ell=3.5$ and a single-mode fiber, into LG modes.  (a) The contribution per
LG mode for mode orders $m+n \le 7$, where $l=n-m$ and
$p=\min(n,m)$. (b) The fidelity of the mode decomposition as a function of
number of modes used in the decomposition. Each of the data points
is calculated for LG modes of mode order up to and including the value
indicated on the top axis. The solid line is an accurate fit to the numerical
data, allowing extrapolation.}\label{fig:modal}
\end{figure*}
The most convenient implementation of this scheme is by choosing the
non-integer part $\mathcal{L}$ of the step index $\ell$ to be half-integer,
i.e.\ $\mathcal{L}=\frac{1}{2}$; the states with $\alpha=0$ and $\alpha=\pi$
are
then orthogonal (independent of the integer part of $\ell$) \cite{WOE04:3}.
When we place an analyzer containing an SPP with index $\ell$ in the signal
path of a SPDC setup, and one containing an SPP with index $-\ell$ in the
idler path, the coincidence probability as a function of the orientations of
the two SPPs is predicted to be \cite{WOE04:3}
\begin{equation}\label{eq:coincidenceprobability}
P_c(\alpha_s,\alpha_i)=
  C\left(1-\frac{\left|\alpha_s-\alpha_i\right|}{\pi}\right)^2,
\end{equation}
viz.\ a parabolic coincidence fringe with a visibility of unity. The constant
$C$ depends on the overlap of the radial part of the two-photon state as
emanating from the SPDC crystal, with the radial part of the fiber
mode \cite{WOE04:3}. This parabolic coincidence fringe, which is a function of
only the \emph{relative} orientation of the two SPPs, reveals the entanglement
of the twin photons.

The key to success in the experiment lies in the analyzers or, more
specifically, in the two SPPs. Our SPPs have been manufactured using a polymer
molding technique starting from a micro-machined mold, yielding identical
plates with $\ell=\pm 3.48\pm 0.02$ at the (degenerate SPDC) wavelength of 813~nm.
The step of each SPP is not abrupt but in fact a linear ramp with an azimuthal
width of $6^\circ$. Furthermore, the plates show a central `anomaly' with a
diameter of about $300~\mu$m \cite{WOE04:2}. The rest of the plate has an rms
deviation of $\approx15$~nm from the ideal shape \cite{WOE05:rmsfootnote}.

The high-dimensional nature of the $\ell=3.5$ SPP is illustrated in
\figref{modal}(a), which shows a modal decomposition for mode order $\le 7$ if
the input is a fundamental Gaussian LG$^0_0$ (i.e.\ the fiber mode). The
fidelity of this truncated decomposition is only $43\%$ (\figref{modal}(b)).
When
truncating at mode order 20 (corresponding to 231 LG modes), the fidelity
becomes $71\%$. Extrapolation shows that a fidelity of 98\% requires about
125000 LG modes. We have checked that these results remain essentially unchanged
when using the actual topography of our SPP plates; in particular, the finite
azimuthal width of the step has no significant effect.

The SPDC setup is shown in \figref{quantumsetup}(a). The pump laser is a $50$~mW
Krypton-ion laser (Spectra Physics 2060) operating at a wavelength of
$406.7$~nm. Its output is weakly focused on a nonlinear crystal (BBO), with
beam waist $w_0=0.78\pm 0.01$~mm. The crystal has a thickness $L=1.0$~mm and
has been cut for nearly collinear type-II SPDC.  Before inserting a
fractional-OAM analyzer in each arm, we test the setup by observing
polarization entanglement; we achieve 95\% and 92\% fringe visibility in the
crystal (H,V) and $45^\circ$ basis, respectively \cite{KWI95}.

We fix the polarizers in the crystal basis and insert a fractional-OAM
analyzer with SPP index $\ell_s=+3.48$ in the signal path and an analyzer with
SPP index $\ell_i=-3.48$ in the idler path. The SPDC cone-crossings in the far
field are found to be much larger than the far-field intensity pattern of the
analyzer mode, as illustrated in \figref{quantumsetup}(b). The SPDC intensity
varies about $10\%$ over the SPP diffraction profile.

The SPPs have to be aligned with great precision. In particular, both the
rotation axis and the center of each SPP have to be aligned, with a precision
of order $1~\mu$m, with the core of the corresponding single-mode fiber. Only
when this amount of care is invested in aligning the optical elements, can one
rotate both SPPs to make use of the $\alpha_s$ and $\alpha_i$ degrees of
freedom without losing any signal.

\begin{figure}[!th]
\centerline{\includegraphics[width=6.464cm]{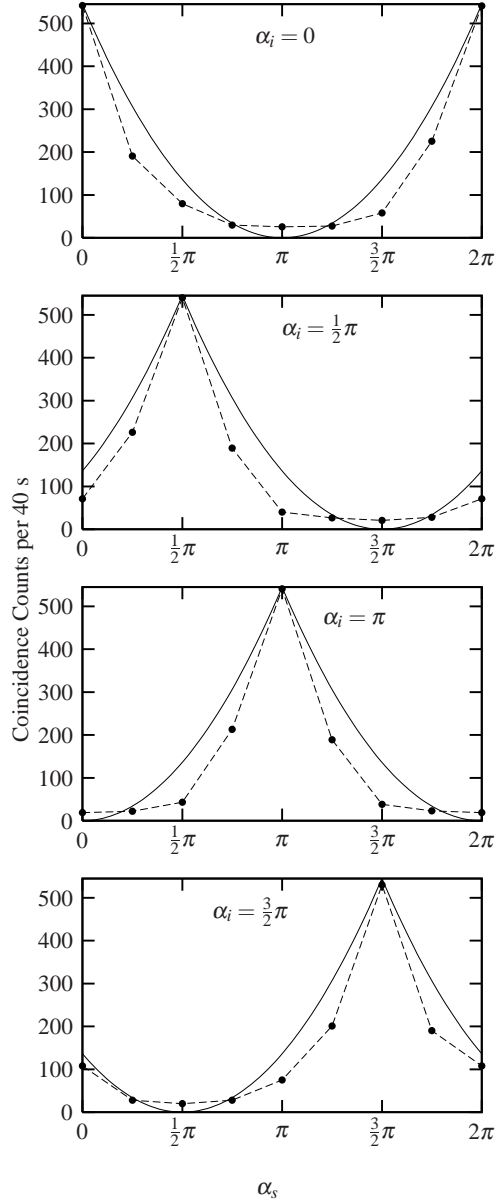}}
\caption{The measured coincidence fringes (circles and broken line segments
to guide the eye) and the theoretical fringes (curves) as a function of the
orientation $\alpha_s$ of the SPP in the signal arm. The parabolic fringe
shifts concordantly with the orientation $\alpha_i$ of the SPP in the idler
arm.}\label{fig:quantumresults}
\end{figure}

Our experimental results are shown in \figref{quantumresults}: four coincidence
fringes (obtained for idler SPP-settings $\alpha_i=0,\pi/2,\pi,3\pi/2$) as a
function of the signal SPP-setting $\alpha_s$. These fringes are shown
together with the expected coincidence fringes as predicted by
\equref{coincidenceprobability}. The data show that the coincidence-count
rate as a function of the orientation of the signal SPP has a parabola-like
shape. Additionally, when the idler SPP is re-oriented, we see that the fringe
shifts accordingly. Consequently, we find, according to the quantum-mechanical
prediction, that the parabolic coincidence fringe depends only on the
\emph{relative} orientation of the SPPs.
Analytical calculations in which one assumes a completely separable (as
opposed to entangled) two-photon
state while OAM is conserved per pair, result in fringes of \emph{zero}
visibility. In the context of a quantum-mechanical description of the SPDC
process, the experimental data show thus entanglement. However, based on the
present set of data, we cannot prove quantum non-locality using the
two-dimensional Clauser-Horne-Shimony-Holt version of the Bell inequality
\cite{HOL69}, in contrast to what we predicted earlier
\cite{WOE04:3,WOE05:bellfootnote}.

The dimensionality of the Hilbert space in which each photon of the pair
lives, must be $D>2$ as can be concluded from the parabolic shape of the
fringes in \figref{quantumresults}; after all, for $D=2$, as in polarization
entanglement, a sinusoidal fringe would appear.  Regretfully, the fringes do
not directly betray the precise dimensionality; nevertheless, we can make an
estimate based on the following considerations.

The dimensionality $D$ of the subspace probed in the experiment is either
limited by the number of entangled modes emitted by the BBO crystal, expressed
by the Schmidt number $K$ \cite{EBE00,TOR03}, or by the (spatial) modal
bandwidth of the analyzers, $N$.  To estimate $K$, we follow Law and Eberly's
approximate analytical theory
\cite{EBE04},
\begin{equation}
K > \frac{1}{4}\left(\sqrt{\frac{L}{w_0^2k_p}} +
\sqrt{\frac{w_0^2k_p}{L}}\right)^2.
\end{equation}
Our experimental values for the pump-beam waist $w_0$, crystal length $L$ and
the pump wavenumber $k_p$ result in a lower limit to the Schmidt number, $K >
3700\pm 100$. A comparison between the approximate and exact theories for the
value of $K$ shows that, for high values of $K$, as in the present case, the
approximate theory yields a severe underestimation.

To make sure that our analyzers are not limiting the dimensionality of the
probed subspace to $D < 3700$, we need to establish that their modal bandwidth
$N\gg 3700$. This can be accomplished by examining the conversion
efficiency of the two analyzers, positioned as an optical train in a
classical-optics experiment: we use one analyzer to
convert the fundamental Gaussian mode from a laser beam to an extended
superposition of LG modes, while the other is used to convert this
superposition back to the fundamental Gaussian mode. The conversion
efficiency, which we found to be 98\%, basically tells us how well the modal
spectra of the two analyzers overlap. Combined with knowledge of the modal
spectrum of a single analyzer (cf.\ \figref{modal}), this value yields the modal
bandwidth $N$ of the analyzers. The experimental value of 98\% \cite{WOE04:2},
which is the fidelity of one analyzer mode as measured by the other analyzer,
corresponds to a modal bandwidth $N\approx 125000$ (which corresponds, in
fact,
to the diffraction limit of the full optical setup).  Consequently, the
analyzers are not a bottleneck in this experiment and we can therefore
conclude that the dimensionality of the probed Hilbert space is $3700 < D <
125000$.

In conclusion, we have reported an experiment on SPDC twin-photons
demonstrating, within the quantum-mechanical description of SPDC, the
entanglement of twin photons with respect to their fractional OAM. As this
experiment involves a unitary transformation of basis from the natural LG
states with integer OAM to a new basis with fractional OAM, the data
explicitly shows that a very large part of the spatial Hilbert space is
entangled, the dimensionality of which we estimate as $D>3700$ per photon.
The entangled space that we have probed in this experiment only represents one
particular cross section of the full Hilbert space of (spatial) two-photon
states.
Presently, we are designing analyzers with alternative spatial geometries
and step heights, and, additionally, analyzers that operate on both
amplitude and phase of the photon state.  Experiments with such analyzers will
possibly allow us to sample a large variety of cross sections of the full
Hilbert space.

\section*{Acknowledgments}
We acknowledge J.~A.~W.~van~Houwelingen for his work on testing the spiral
phase plates. This work is part of the research program of the `Stichting voor
Fundamenteel Onderzoek der Materie (FOM)' and is supported by the EU program
ATESIT.


\begin{thebibliography}{19}
\bibitem[{WOE({\natexlab{a}})}]{WOE05:rmsfootnote}
\bibinfo{note}{This was established with a Zygo interferometer.}
\bibitem[{WOE({\natexlab{b}})}]{WOE05:bellfootnote}
\bibinfo{note}{Although the orhogonal half-integer $\ell$ states with
  $\alpha=0$ and $\alpha=\pi$ may be used to span a two-dimensional subspace
  \cite{WOE04:3}, continuously varying $\alpha$ spans a space of much larger
  dimensionality, invalidating the dichotomy required by the fair-sampling
  assumption \cite{RAP81} in the smaller, 2D subspace, leading to an invalid
  Bell parameter. Instead, as alluded to earlier \cite{WOE04:3}, a generalized
  Bell inequality has to be used \cite{POP02,ZEI02:2}.}
\end{thebibliography}

\end{document}